\documentclass[twocolumn,showpacs,preprintnumbers,amsmath,amssymb]{revtex4}

\usepackage{epsfig}
\usepackage{graphicx}
\usepackage{dcolumn}
\usepackage{bm}

\newcommand{\oo}{{\omega_\oplus}}
\newcommand{\om}{\omega}

\newcommand{\ot}{\omega_t}

\newcommand{\Oo}{\Omega_\oplus}

\newcommand{\be}{\begin{equation}}
\newcommand{\ee}{\end{equation}}
\newcommand{\bo}{\beta_\oplus}

\begin{document}

\preprint{APS/123-QED}

\title{Relativity tests by complementary rotating Michelson-Morley experiments}

\author{Holger M\"uller}
\email{holgerm@stanford.edu} \affiliation{Physics Department,
Stanford University, 382 Via Pueblo Mall, Stanford, California
94305, USA}
\author{Paul Louis Stanwix, Michael Edmund Tobar, Eugene Ivanov}
\affiliation{University of Western Australia, School of Physics
M013, 35 Stirling Hwy., Crawley 6009 WA, Australia}
\author{Peter Wolf}
\affiliation{LNE-SYRTE, Observatoire de Paris, 61 Av. de
l'Observatoire, 75014 Paris, France}
\author{Sven Herrmann, Alexander Senger, Evgeny Kovalchuk, and Achim Peters}
\affiliation{Institut f\"ur Physik, Humboldt-Universit\"at zu
Berlin, Hausvogteiplatz 5-7, 10117 Berlin, Germany}
\date{\today}

\begin{abstract}
We report Relativity tests based on data from two simultaneous
Michelson-Morley experiments, spanning a period of more than one
year. Both were actively rotated on turntables. One (in Berlin,
Germany) uses optical Fabry-Perot resonators made of fused silica;
the other (in Perth, Australia) uses microwave whispering-gallery
sapphire resonators. Within the standard model extension, we
obtain simultaneous limits on Lorentz violation for electrons (5
coefficients) and photons (8) at levels down to $10^{-16}$,
improved by factors between 3 and 50 compared to previous work.
\end{abstract}

\pacs{Valid PACS appear here}
\maketitle



The original Michelson-Morley (MM) experiment provided physicists
with a first glimpse of Lorentz invariance when relativistic
velocities were inaccessible to experiments and the theory of
relativity was yet to be formulated. In a similar way, modern
versions of outstanding precision can attempt to detect minuscule
{\em violations} of Lorentz invariance and thus provide physicists
with a first glimpse of effects of a future theory of quantum
gravity in the low-energy limit
\cite{Will93,Amelino-Cameliaetal05,Mattingly05}. Since the form of
the putative violations is not predetermined, it is important to
probe a broad variety of them experimentally.

In contrast to the original interferometer experiments \cite{MM},
modern MM-experiments are generally based on a measurement of the
resonance frequencies \be\label{fundamental}
\omega=2\pi\frac{mc}{2nL} \ee ($m$ is a constant mode number, $c$
the velocity of light, and $n$ the index of refraction if a medium
is present) of standing waves in resonant optical or microwave
cavities. Any type of Lorentz violation that affects the isotropy
of $c$ \cite{KosteleckyMewes}, $L$ \cite{TTSME,ResSME,TLVMC}, or
$n$ \cite{TLVMC} can potentially be detected. $L$ and $n$ are
properties of macroscopic matter and thus sensitive to Lorentz
violation in the Maxwell and Dirac equations that govern its
constituents. However, each of the simple MM-experiments performed
so far (recently,
\cite{BrilletHall,cavitytests,MuellerMM,Antonini,Wolfsme,MMSven,Stanwix,Stanwixnew,Lipa})
does not by itself provide enough information to distinguish
between the different influences and thus can only bound
combinations of them. To remove these restrictions, experiments
featuring dissimilar cavities that have a different dependence of
$L$ and $n$ on Lorentz violation have been suggested
\cite{ResSME,TLVMC}.

\begin{figure}
\centering
\epsfig{file=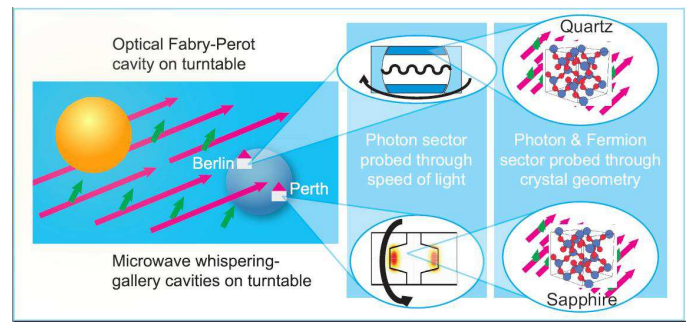,
width=0.5\textwidth} \caption{\label{concept} Modifications of $c,
n,$ and $L$ are probed for cavities of different geometry and
material, which allows us to separately probe Lorentz violation in
the electron and photon sector (symbolized by the arrows).}
\end{figure}

Here, we report on the first realization of such simultaneous,
complementary MM-experiments that use different cavity materials,
geometries, operating frequencies, and locations (Berlin, Germany
and Perth, Australia) in different hemispheres (Fig.
\ref{concept}). Both provide data spanning a period of more than
one year and were performed on a rotating table. This allows us to
use Earth's orbital motion and rotation, as well as the
turntables, to modulate the `laboratory' frame of reference and
thus to restrict Lorentz violations of more general symmetries
than otherwise possible. By combining all data, we obtain
independent, simultaneous limits on a broader range of Lorentz
violations in the Dirac and Maxwell sectors than any single
experiment could.

We use the standard model extension (SME) \cite{SME} as a
comprehensive framework for Lorentz violation. It extends the
Lagrangians of the standard model by the most general Lorentz
violating terms that can be formed from the standard model fields
and Lorentz tensors. Modifications of the photon sector
\cite{Ni,KosteleckyMewes} are described by a tensor
$(k_F)_{\kappa\lambda\mu\nu}$ entering the Maxwell equations. Ten
of its elements lead to a dependence of $c$ on the polarization.
The observation that the apparent polarization of certain
astronomical sources does not depend on the wavelength bounds them
to levels between $10^{-37}$ and $10^{-32}$
\cite{KosteleckyMewes,Amelino-Cameliaetal05}. The remaining nine
elements of $(k_F)_{\kappa\lambda\mu\nu}$ can only be measured in
laboratory instruments that probe the isotropy of $c$. They can be
arranged into traceless $3\times 3$ matrices $\tilde\kappa_{e-}$
(symmetric) and $\tilde\kappa_{o+}$ (antisymmetric)
\cite{KosteleckyMewes}. $\tilde\kappa_{e-}$ and
$\tilde\kappa_{o+}$ also affect $L$ and $n$ due to a modification
of the Coulomb potential, but this has been shown to be negligible
for most experiments, including ours \cite{TTSME}.

Non-negligible shifts in $L$ and $n$, however, result from Lorentz
violation in the electron sector \cite{ResSME,TLVMC,H2SME}. In the
non-relativistic limit, a modification of the electron's
energy-momentum relation according to $p^2/(2m)\rightarrow
(p^2+p_jp_k E^{jk})/(2m)$ where $p$ is the 3-momentum and
$E'_{jk}=-c_{jk}-\frac12 c_{00}\delta_{jk}$ is given by a SME
tensor $c_{\mu\nu}$ that enters the Dirac Lagrangian of the free
electron. The resulting modification of the electronic states
within solids leads to a change of $L$ that is given by the
diagonal elements of a strain tensor
\begin{equation}
e_{jk}=\frac 12 \left(\frac{\partial L_j}{\partial
x_k}+\frac{\partial L_k}{\partial x_j}\right)=\mathcal B_{jklm}
E'_{lm}\,.
\end{equation}
The sensitivity tensor $\mathcal B$ is predicted in detail by
invoking perturbation theory for the electrons as described by
Bloch wave functions \cite{ResSME,TLVMC}. This theory also
predicts a change of the index of refraction. For materials of
trigonal or higher symmetry and microwave frequencies,
\begin{equation}
\frac{\delta n}{n}=\bar\beta E'_3, \quad
\bar\beta=\frac{(n^2-1)(n^2+2)}{3n^2}(\mathcal B_{31}-\mathcal
B_{13})
\end{equation}
and $E'_3=E'_{11}+E'_{22}-2E'_{33}$
\cite{TLVMC}. Because of the material dependence of $\mathcal B$
and $\bar \beta$, experiments using cavities of different nature
measure independent combinations of the elements of $k_F$ and
$c_{\mu\nu}$.

Throughout this work, we use the conventions made in Refs.
\cite{ResSME,TLVMC}. In particular, by definition of coordinates
and fields, we take the proton sector to be Lorentz invariant.
This is always possible \cite{SME,KosteleckyMewes} and leads to an
unambiguous definition of the $c$- and $k_F$- coefficients.


If Lorentz invariance is violated, the resonance frequency
$\omega$ of our cavities will exhibit a measurable modulation
having Fourier components at frequencies that are integer
combinations of twice the angular frequency of the turntable
$2\om_t$, Earth's rotation $\oo$, and Earth's orbit $\Oo$. Such a
signal can be expressed as $\delta\nu/\bar\nu=B\sin2\ot T+C\cos
2\ot T$, where
\begin{eqnarray} \label{BC} B=B_{0}+B_{s1}\sin\oo T+B_{c1}\cos\oo
T +B_{s2}\sin 2\oo T\nonumber \\ +B_{c2}\cos2\oo T, \quad
C=C_{0}+C_{s1}\sin\oo T\nonumber
\\ +C_{c1}\cos\oo T+C_{s2}\sin2\oo T+C_{c2}\cos2\oo T
\end{eqnarray}
are themselves time-dependent. This is a general expression,
applicable in any test model that leads to modulation at (some of)
these frequencies. In the SME, the $B, C$ will depend on different
linear combinations of the elements of $k_F$ and $c$. By analyzing
the measured cavity frequency in terms of the above modulations,
separate measurement of the tensor elements is thus possible. An
experiment without turntable can measure four combinations. If at
least 1\,y of data is taken, three additional ones, that depend on
the Lorentz boost due to Earth's orbit and cause modulation
components differing by $\Oo=2\pi/1$\,y in frequency, can be
resolved. Finally, use of a turntable provides access to an eighth
coefficient, which is otherwise suppressed due to the presence of
a symmetry axis (Earth's axis). However, to separately measure the
changes in $L$ and $n$ is only possible with complementary
experiments that use different cavities and materials.


The Berlin setup compares the resonance frequencies of two
monolithic, diode-pumped neodymium:YAG lasers at a wavelength of
1064\,nm that are stabilized to resonances of Fabry-Perot cavities
fabricated from fused silica (with BK7 substrates for the
mirrors). One (L=2.85\,cm, Finesse $\mathcal F=1.7\times 10^5$) is
continuously rotated at a rate of $1/(43$\,s). To reduce
systematic effects associated with table rotation, a precision air
bearing turntable (type RTV 600, Kugler GmbH, Salem, Germany) is
used that is specified for $<0.1\,\mu$rad rotation axis wobble.
The rotation axis is actively stabilized to the vertical using a
tilt sensor (Applied Geomechanics, Inc.) on the rotating platform.
The other cavity (L=10\,cm, $\mathcal F=2\times 10^4$) is oriented
north-south.

\begin{figure}
\centering
\epsfig{file=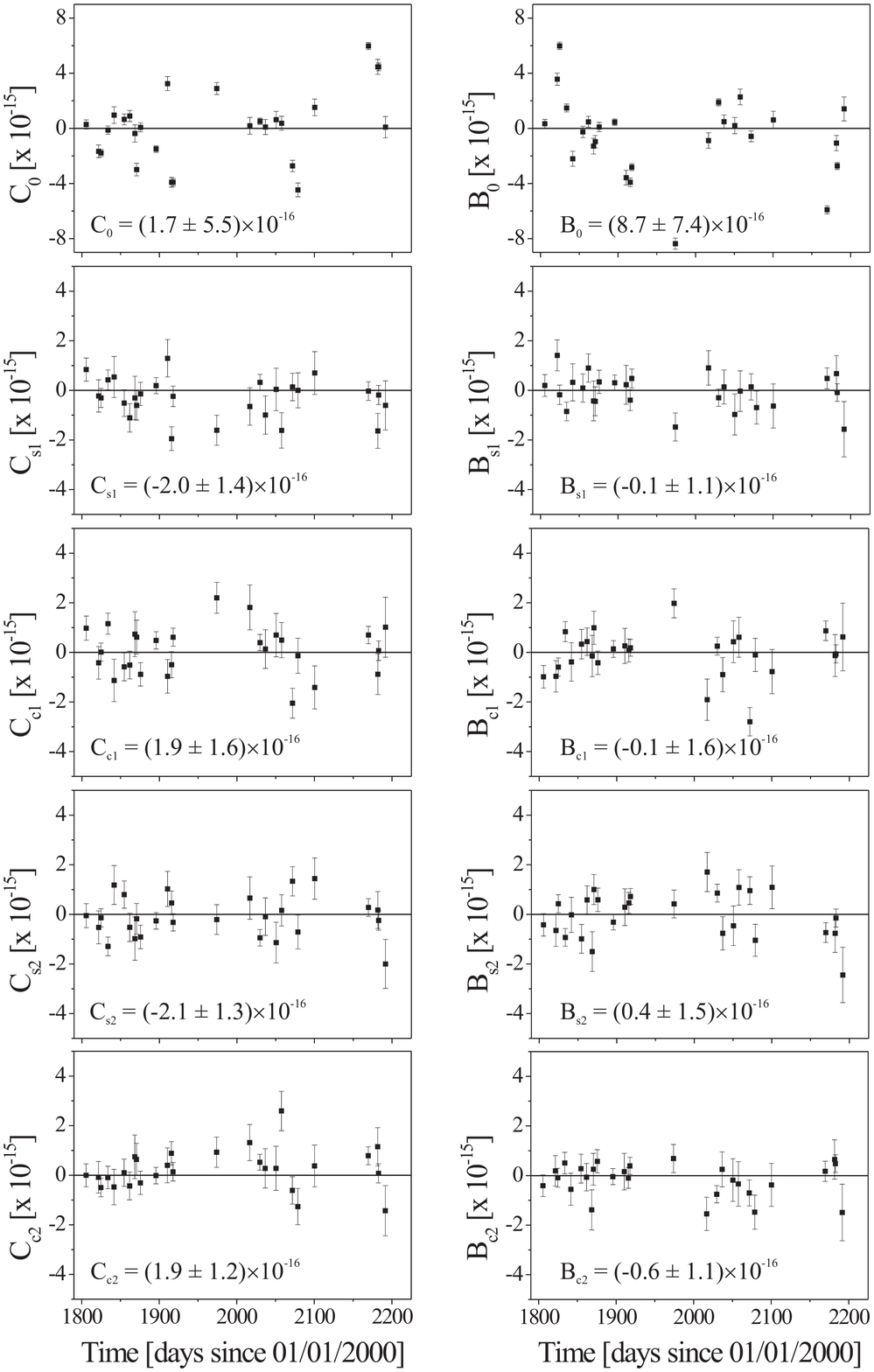,width=0.5\textwidth}
\caption{\label{results} $B_k, C_k$ versus time.}
\end{figure}

Data has been collected for 396\,d, totalling to 62\,d of useful
data in 27 sets (118,000 turntable rotations), beyond the data
already reported in \cite{MMSven}. To analyze the data, we first
break it down into $i=1....N$ subsets of 10 table rotations each.
This approaches an optimal filter that rejects possible signal
components due to drift of the cavity frequency while passing the
sinusoidal signals. The subsets are individually fitted with the
sine and cosine amplitudes $B(t_i)$ and $C(t_i)$ (each is taken as
constant over the subsets and assigned the subset's mean time
$t_i$). Systematic influences are allowed for in the fit function
by including sine and cosine amplitudes at $\om_r$, a constant
offset, and drift terms linear and quadratic in time. This yields
192 values for each of $B(t_i)$ and $C(t_i)$ from each 24 hours of
data. Then, the $B(t_i), C(t_i)$ coefficients are fitted with Eq.
(\ref{BC}). Performing similar fits on all 27 data sets yields one
set of
$B_k$ and $C_k$ from each, see Fig. \ref{results}. 

Residual systematic effects at $2\om_r$ primarily affect the
coefficients $C_0$ and $B_0$ at a level of $5\sigma$ within
individual subsets. They differ in phase and magnitude (see Fig.
\ref{results}) and average out in the final result. This, however,
leaves an increased error bar on the $C_0$ and $B_0$ averages. The
other amplitudes have relatively smaller error bars, as they are
are affected by systematics only indirectly through additional
time-dependent influences, such as a daily modulation of
temperature or tilt of the building floor.


The hypothetical signal for Lorentz violation in the SME,
calculated from the motion of the cavities by following the
methods described in Refs. \cite{KosteleckyMewes,TTSME,TLVMC} is
detailed in Tab. \ref{Berlincoeffs}. As throughout the literature,
the coefficients for Lorentz violation with capital indices are
referred to a sun-centered celestial equatorial reference frame
defined, for example, in Ref. \cite{KosteleckyMewes}. To estimate
the SME-coefficients, we fit this hypothetical signal for Lorentz
violation to the experimental results shown in Fig. \ref{results},
weighted according to the inverse squared fit error. The results
for the $\lambda^{IJ}\equiv (\tilde\kappa_{e-})^{IJ}+2\mathcal B_q
c_{IJ}$ (where $2 \mathcal B_q=-5.92$ is a material parameter for
fused quartz \cite{TLVMC}) are, in parts in $10^{16}$,
\begin{eqnarray}
\lambda^{XY}&=&-4.9(2.5), \quad \lambda^{XZ}=-1.4(2.5),\nonumber\\
\lambda^{YZ}&=&4.1(3.9), \quad \lambda^{ZZ}=13.3(9.8),\nonumber
\\ \lambda^{XX}-\lambda^{YY}&=&05.7(22.6)\,.
\end{eqnarray}
The results for $(\tilde\kappa_{o+})$ are, in parts in $10^{12}$,
\begin{eqnarray}
(\tilde\kappa_{o+})^{XY}&=&5.7(3.7),\quad
(\tilde\kappa_{o+})^{XZ}=5.3(6.3),\nonumber\\
(\tilde\kappa_{o+})^{YZ}&=&-0.2(6.2).
\end{eqnarray}

\begin{table*}
\centering \caption{\label{Berlincoeffs} Signal components for the
Berlin setup. $\lambda^2\equiv \lambda^{XX}-\lambda^{YY}$.}
\begin{tabular}{cccc}\hline\hline
$C_{0}$ & $\frac14 \sin^2\chi(\frac32
\lambda^{ZZ}-\bo[(\cos\eta\kappa_{o+}^{XZ}+2\kappa_{o+}^{XY}\sin\eta)\cos\Oo
T'+\kappa_{o+}^{YZ}\sin\Oo T']$ & $B_{0}$ & 0 \\ $C_{s1}$ &
$\frac12\cos\chi\sin\chi(-\lambda^{YZ}+\bo[\kappa_{o+}^{XY}\cos\eta-\kappa_{o+}^{XZ}\sin\eta]\cos\Oo
T'$ & $B_{s1}$ & $-C_{c1}/\cos\chi$ \\
$C_{c1}$ & $\frac12\cos\chi\sin\chi(-\lambda^{XZ}+
\bo[\kappa_{o+}^{YZ}\sin\eta\cos\Oo T'-\kappa_{o+}^{XY}\sin\Oo
T']$ & $B_{c1}$ & $C_{s1}/\cos\chi$  \\ $C_{s2}$ &
$\frac14(1+\cos^2\chi)(\lambda^{XY}-
\bo[\kappa_{o+}^{YZ}\cos\eta\cos\Oo T'+\kappa_{o+}^{XZ}
\sin\Oo T']$ & $B_{s2}$ & $-2\cos\chi/(1+\cos^2\chi)C_{c2}$ \\
$C_{c2}$ & $\frac14(1+\cos^2\chi)(\frac12\lambda^2-
\bo[\kappa_{o+}^{XZ}\cos\eta\cos\Oo T'-\kappa_{o+}^{YZ}\sin\Oo
T'])$ & $S_{c2}$ & $2\cos\chi/(1+\cos^2\chi)C_{s2}$ \\
\hline\hline
\end{tabular}
\end{table*}


The Perth setup compares the frequencies of two orthogonally
orientated high Q-factor ($2\times 10^8$) cryogenically cooled
($\sim 6$\,K) microwave resonators. Each consists of a sapphire
crystal mounted inside a metallic shield. The crystal is 3\,cm
diameter and height, and is machined with its crystal axis in line
with the cylindrical axis. Each resonator is excited in the
whispering gallery WGH$_{8,0,0}$ mode at approximately 10\,GHz by
two separate Pound stabilized oscillator circuits, with a
difference frequency of about 226\,kHz. The WGH$_{8,0,0}$ mode has
dominant electric and magnetic fields in the axial and radial
directions respectively, corresponding to a Poynting vector around
the circumference. 98\% of the electromagnetic energy is confined
within the sapphire crystal so it is subject to the index of
refraction of sapphire ($n_\perp=3.04, n_\parallel=3.37$ at 6\,K).
The two resonators are oriented with their cylindrical axis
perpendicular to each other in the horizontal plane and placed
inside a vacuum chamber and cryogenic dewar, which is mounted in a
turntable and rotated at 1/(18\,s) about its vertical axis.

As discussed in a previous publication \cite{Stanwixnew}, the data
used in this analysis spans a period from December 2004 to January
2006. It consists of 27 sets of data totalling approximately 121
days. The data analysis proceeds in analogy to the Berlin setup.
The data is broken down into blocks of 40 rotations, each of which
is fitted with $B(t_i)$ and $C(t_i)$. The hypothetical signal for
Lorentz violation in the photon sector is listed in Tab. I of
\cite{Stanwixnew}. We calculate the effects of the Dirac sector as
in \cite{TTSME,TLVMC}, taking into account both the changes in the
geometry of the cavity and the index of refraction. We find that
the combined signal can be expressed by substituting
$(\tilde\kappa_{e-})^{IJ}\rightarrow \mu^{IJ}=
(\tilde\kappa_{e-})^{IJ}+3\mathcal B_{\rm s} c_{IJ}$ in that
table, where $3\mathcal B_{\rm s}\equiv 3[-(\frac 13-\frac 12
\mathcal B_{13})+\bar\beta] \approx -2.25$ is a material and
geometry-dependent coefficient for the sapphire WGR \cite{TLVMC}.
Fitting the hypothetical signal to the data leads to (parts in
$10^{16}$)
\begin{eqnarray}
\mu^{XY}&=&2.9(2.3),\quad \mu^{XZ}=-6.9(2.2),\nonumber \\
\mu^{YZ}&=&2.1(2.1),\quad \mu^{ZZ}=143(179),\nonumber
\\ \mu^{XX}-\mu^{YY}&=&-5.0(4.7),
\end{eqnarray}
and (parts in $10^{12}$)
\begin{eqnarray}
(\kappa_{o+})^{XY}&=&-0.9(2.6),\quad
(\kappa_{o+})^{XZ}=-4.4(2.5),\nonumber \\
(\kappa_{o+})^{YZ}&=&-3.2(2.3).
\end{eqnarray}

\begin{table}[t]
\centering \caption{\label{HerrmannStanwix} Results on electron
and photon coefficients $\tilde\kappa_{e-}$ and $c$ in units of
$10^{-16}$ and $\tilde\kappa_{o+}$ in $10^{-12}$ (one sigma
errors).}
\begin{tabular}{ccccc} \hline
$\tilde\kappa_{e-}^{XX}-\tilde\kappa_{e-}^{YY}$ &
$\tilde\kappa_{e-}^{XY}$ & $\tilde\kappa_{e-}^{XZ}$
&$\tilde\kappa_{e-}^{YZ}$ & $\tilde\kappa_{e-}^{ZZ}$ \\
$-12(16)$  & $7.7(4.0) $& $-10.3(3.9)$&$0.9(4.2)$ & $223(290)$ \\
\hline
$c_{XX}-c_{YY}$& $c_{(XY)}$ &$c_{(XZ)}$ & $c_{(YZ)}$& $c_{3}$ \\
$-2.9(6.3)$ & $2.1(0.9)$ &$-1.5(0.9)$ & $-0.5(1.2)$ &$-106(147)$
\\ \hline
$\lambda^{ZZ}$ & $\tilde\kappa_{o+}^{XY}$ & $\tilde\kappa_{o+}^{XZ}$ & $\tilde\kappa_{o+}^{YZ}$ \\
$13.3(9.8)$ & 1.7(2.0) & -3.1(2.3) & -2.8(2.2)  \\
 \hline
\end{tabular}
\end{table}


The data from both setups are of similar quality. However, the
constraints on the $\tilde\kappa_{o+}$ from Perth have $\sim 2$
times lower confidence interval. On the other hand, $\lambda^{ZZ}$
from the Berlin setup is about $17$ times more accurate than
$\mu^{ZZ}$ from Perth, as this signal occurs at $2\om_r$, at which
frequency signals from wobble of the turntable have a strong
Fourier component. The precision turntable along with an active
vertical alignment of the rotation axis leads to this higher
accuracy.

Combined, the constraints from both setups, shown in Eqs. (3-6),
are sufficient to calculate separate bounds on Lorentz violation
in the electron and photon sector, see Tab. \ref{HerrmannStanwix}.
We also included the limit on $\lambda^{ZZ}$ from the Berlin
setup. For the elements of $\kappa_{o+}$, the two experiments
provide complementary limits. The ones listed in the table are
obtained from both by weighted averaging. We note that
$\tilde\kappa_{e-}^{XZ}$ and $c_{XY}$ are at the (2-3)$\sigma$
level. However, systematic effects associated with rotation of the
turntable are extremely hard to quantify at this level, since with
the given noise of the data this takes a full year of averaging.
To enable a better characterization of the systematics or of a
possible signal, lower noise is required. This will be achieved in
the next generation of rotating experiments with cavities having
even higher quality factors, which promise to reduce the noise by
more than one order of magnitude. Therefore, for the time being,
we regard our results as a confirmation of Lorentz Invariance. In
the future, birefringence- \cite{TLVMC} or dual-mode- \cite{Tobar}
cavity experiments can overcome some of the systematic effects.


In summary, we present relativity tests based on simultaneous,
complementary Michelson-Morley experiments. Use of dissimilar
cavities, operation on both hemispheres, and extensive data-taking
over a period of $>1\,$y makes this the first simultaneous
measurement of a complete set of limits on spin-independent
Lorentz violation in the electron and photon sectors. We determine
14 limits on Lorentz violation parameters of the standard model
extension. Compared to the best previous limits that do not use
assumptions on the vanishing of Lorentz violation in one sector
\cite{MuellerMM,Wolfsme,TLVMC,Baltschu}, they are improved by
factors between $\sim 3-50$. Thus, we also confirm the isotropy of
the velocity of light without using such assumptions.


We would like to thank S. Chu and G. Ertl for support and
important discussions. This work was supported by the Australian
Research Council. H.M. thanks the Alexander von Humboldt
Foundation and S.H. the Studienstiftung des Deutschen Volkes.


\begin{references}

\bibitem{Will93} C.M. Will, {\sl Theory and experiment in gravitational physics
(revised edition)}, (University Press, Cambridge, 1993)

\bibitem{Amelino-Cameliaetal05} G. Amelino-Camelia,
C. L\"ammerzahl, A. Macias, and H. M\"uller in: A.Macias, C.
L\"ammerzahl, and D. Nunez (eds.), Gravitation and Cosmology, page
30. AIP Conference Proceedings {\bf 758}, Melville, N.Y. (2005).

\bibitem{Mattingly05} D. Mattingly, Living Reviews {\bf 8,} http://www.livingreviews.org/lrr–2005–5,
2005 (cited on April 4, 2006).

\bibitem{MM} A.A. Michelson, Am. J. Sci. {\bf 22}, 120 (1881);
A.A. Michelson and E.W. Morley, Am. J. Sci {\bf 34}, 333 (1887);
Phil. Mag. {\bf 24}, 449 (1897)

\bibitem{KosteleckyMewes} V.A. Kosteleck\'{y} and M. Mewes, Phys. Rev. D {\bf 66}, 056005
(2002); \prl {\bf 97,} 140401 (2006).

\bibitem{TTSME} H. M\"uller, C. Braxmaier, S. Herrmann, A. Peters, and C. L\"ammerzahl,
Phys. Rev. D {\bf 67,} 056006 (2003).

\bibitem{ResSME} H. M\"{u}ller, S. Herrmann, A. Saenz, A. Peters, and C. L\"ammerzahl,
Phys. Rev. D {\bf 68,} 116006 (2003).

\bibitem{TLVMC} H. M\"uller, Phys. Rev. D {\bf 71,} 045004 (2005).

\bibitem{BrilletHall} A. Brillet and J.L. Hall, Phys. Rev. Lett. {\bf 42,} 549 (1979).

\bibitem{cavitytests} D. Hils and J.L. Hall, Phys. Rev. Lett. {\bf 64,} 1697 (1990);
C. Braxmaier {\em et al., ibid.} {\bf 88,} 010401 (2001); H.
M\"uller {\em et al.,} Int. J. Mod. Phys. D {\bf 11}, 1101 (2002);
P. Wolf {\em et al.,} Phys. Rev. Lett. {\bf 90}, 060402 (2003).

\bibitem{MuellerMM} C. Braxmaier, H. M\"{u}ller, O. Pradl, J.
Mlynek, A. Peters, and S. Schiller, Phys. Rev. Lett. {\bf 88,}
010401 (2002); H. M\"uller, S. Herrmann, C. Braxmaier, S.
Schiller, and A. Peters, {\em ibid.} {\bf 91,} 020401 (2003);
Appl. Phys. B (laser opt.) {\bf 77,} 719 (2003); H. M\"uller {\em
et al.,} Opt. Lett. {\bf 28,} 2186 (2003).

\bibitem{Wolfsme} P. Wolf {\em et al.,} Gen. Relativ. Gravitat. {\bf 36,} 2351 (2004);
Phys. Rev. D {\bf 70}, 051902(R) (2004).

\bibitem{Antonini} P. Antonini {\em et al.}, Phys. Rev. A {\bf 71,}
(R), 050101 (2005).

\bibitem{MMSven} S. Herrmann, A. Senger, E. Kovalchuk, H. M\"uller, and A. Peters,
Phys. Rev. Lett. {\bf 95,} 150401 (2005).

\bibitem{Stanwix} P.L. Stanwix {\em et al.,} Phys.
Rev. Lett. {\bf 95,} 040404 (2005).

\bibitem{Stanwixnew} P.L. Stanwix, M.E. Tobar, P. Wolf, C.R. Locke,
and E.N. Ivanov, \prd {\bf 74,} 081101(R) (2006).


\bibitem{Lipa} J.A. Lipa, J.A. Nissen, S. Wang, D.A. Stricker, and D. Avaloff,
Phys. Rev. Lett. {\bf 90,} 060403 (2003).

\bibitem{SME} D. Colladay and V.A. Kosteleck\'{y}, Phys. Rev. D {\bf 55}, 6760 (1997);
Phys. Rev. D {\bf 58}, 116002 (1998); V.A. Kosteleck\'y, Phys.
Rev. D 69, 105009 (2004).

\bibitem{Ni} W.-T. Ni, Phys. Rev. Lett. {\bf 38,} 301 (1977).

\bibitem{H2SME} H. M\"uller, S. Herrmann, A. Saenz, A. Peters, and C. L\"ammerzahl,
Phys. Rev D {\bf 70,} 076004 (2004).

\bibitem{Baltschu} B. Altschul, Phys.
Rev. Lett. {\bf 96,}, 201101 (2006).


\bibitem{Tobar} M.E. Tobar {\em et al.,} Springer Lect. Notes
Phys. {\bf 702} 416-450 (2006).


\end{references}
\end{document}